\documentclass[twocolumn, superscriptaddress,nofootinbib,floatfix, amsfonts, aps]{revtex4-1}
\usepackage{setspace,amsmath,bm,amssymb,graphicx,mathrsfs,footmisc,multirow,booktabs, ctable}
\usepackage{xcolor, color, framed, ulem}

\begin{document}
%\begin{spacing}{1.0}

\title{X(3872) Production in Relativistic Heavy-Ion Collisions }

\author{Baoyi Chen}\email{baoyi.chen@tju.edu.cn}
\affiliation{Department of Physics, Tianjin University, Tianjin 300350, China}

\author{Liu Jiang}\email{jiangliu\_1997@tju.edu.cn}
\affiliation{Department of Physics, Tianjin University, Tianjin 300350, China}

\author{Xiao-Hai Liu}\email{xiaohai.liu@tju.edu.cn}
\affiliation{Department of Physics, Tianjin University, Tianjin 300350, China}

\author{Yunpeng Liu}\email{yunpeng.liu@tju.edu.cn}
\affiliation{Department of Physics, Tianjin University, Tianjin 300350, China}

\author{Jiaxing Zhao}\email{zhao-jx15@tsinghua.org.cn}
\affiliation{Department of Physics, Tsinghua University, Beijing 100086, China}

\date{\today}
\begin{abstract}
Heavy ion collisions provide a unique opportunity to study the nature of X(3872)
compared with electron-positron and proton-proton (antiproton) collisions.
We investigate the centrality and momentum dependence of X(3872)
in heavy-ion collisions
via the Langevin equation and instant coalescence model (LICM).
When X(3872) is treated as a compact tetraquark state, the tetraquarks are 
produced via the
coalescence of heavy and light quarks near the quantum chromodynamic (QCD) phase transition due to the
restoration of the heavy quark potential at $T\rightarrow T_c$. In the molecular scenario,
loosely bound X(3872) is produced via the coalescence of
$D^0$-$\bar D^{*0}$ mesons in a hadronic medium after kinetic freeze-out.
We employ the LICM to explain both $D^0$ and $J/\psi$ production as a
benchmark. Then we give predictions regarding X(3872) production and the nuclear modification 
factor $R_{AA}^{X(3872)}$. 
We find that the total yield of tetraquark is several times larger than the
molecular production in Pb-Pb collisions.
Although the geometric size of the hadronic molecule is huge,
the coalescence probability is small due to strict constraints on the 
relative momentum between
$D^0$ and $\bar D^{*0}$ in the molecular Wigner function,
which significantly suppresses the molecular yield.

\end{abstract}
%\pacs{ }
\maketitle

\section{Introduction}
\label{sec:introduction}

Since the discovery of $X(3872)$ resonance by Belle in 2003~\cite{Choi:2003ue}, 
their properties of X(3872)
have been extensively studied both experimentally and theoretically~\cite{Acosta:2003zx, Aubert:2005zh,Bignamini:2009sk,Brambilla:2019esw}.
As the mass of X(3872) is just below the ${D}^0\bar D^{*0}$ (or ${\bar D}^0D^{*0}$ 
threshold, it might be
a meson-meson molecular state with very small binding energy
~\cite{mole:first,Close:2003sg,Voloshin:2003nt,Wong:2003xk,Liu:2008tn}.
On the other hand,
its constituent quark content is generally believed to be of $c\bar cq\bar q$ type. 
A superposition of two configurations has also been proposed~\cite{Coito:2012vf,Yamaguchi:2019vea}.
Whether $X(3872)$ is a loosely bound hadronic molecule, a compact tetraquark or just a 
kinematic effect such as triangle singularity 
is still under debate
~\cite{Guo:2017jvc,Esposito:2016noz,Nakamura:2019nwd}.
In proton-proton (pp) collisions,
the multiplicity dependence of the yield ratio $X(3872)/\psi(2S)$
at LHCb~\cite{LHCb:2020sey} has been measured,
and it seems to disfavor the molecular interpretation of
X(3872)~\cite{Esposito:2020ywk,Braaten:2020iqw}.
In Pb-Pb (AA) collisions at the Large Hadron Collider (LHC)~\cite{CMS:2021znk},
the yield ratio $X(3872)/\psi(2S)$ is clearly
greater than that
of pp collisions. These experimental studies necessitate more detailed theoretical
studies about
the internal structure of X(3872).
Heavy ion collisions have provided a unique opportunity.
With the generation of deconfined matter called ``quark-gluon plasma'' (QGP) in
heavy-ion collisions, most primordially produced X(3872) is melted in QGP
due to the strong color screening effect and parton inelastic scatterings 
from the thermal partons.
Additionally, the abundant
number of charm pairs produced in the early stage 
can combine with heavy/light quarks to form new hadrons
at the quantum chromodynamic (QCD) phase transition. This contribution is significantly enhanced when
charm quark densities become large in QGP. Both theoretical and experimental
studies have suggested that this coalescence process becomes essential in
the charmonium and multi-charm hadron production at the LHC~\cite{BraunMunzinger:2000px,Thews:2000rj,Grandchamp:2001pf,Andronic:2003zv,
Yan:2006ve,Chen:2018kfo,Zhao:2016ccp,He:2014tga}.
The coalescence model
and statistical hadronization model 
have been extended to study the hadron production 
in heavy-ion collisions~\cite{Cho:2010db,Cho:2011ew,
Fontoura:2019opw,Zhang:2020dwn,Andronic:2019wva, Esposito:2015fsa}.

In this work, we employ the Langevin equation to simulate the Brownian motion
of charm quarks in QGP and D mesons in a hadronic medium.
After the diffusion,
heavy and light quarks may combine to form D mesons, charmonia, or
tetraquarks ($c\bar c q\bar q$) in QGP. In a hadronic medium, D mesons may also
combine to form a hadronic molecule ($D^0\bar D^{*0}$).
The formation processes are described with the instantaneous coalescence
model (ICM). The final production mainly depends on two
factors: the charm quark spatial and momentum distributions in the medium,
and the Wigner functions
of the formed particle. The Wigner function can be determined
via the Weyl transform of the wave function.
Due to the significant difference
between the geometric sizes of a compact tetraquark
(mean radius $\langle r\rangle_{X}\sim0.3-0.5\ \mathrm{fm}$)~\cite{Esposito:2020ywk}
and a loosely bound hadronic molecule (mean radius
$\langle r\rangle_{X}\sim2.5-22\ \mathrm{fm}$,
see table~\ref{tab-mole})
~\cite{Liu:2008tn,Zhang:2020dwn}, their 
Wigner functions are significantly different from each other in the two scenarios.
Their production is expected to become distinguishable in the two scenarios,
which allows us to study whether the X(3872) wave function is
close to a molecular state or a tetraquark state in heavy-ion collisions.

This paper is organized as follows. In Section (\ref{sec:charm}), we introduce the Langevin
equations for the Brownian motions of heavy quarks and D mesons in the medium.
Heavy quark distributions in phase space are also presented
with different degrees of
momentum thermalization.
In Section (\ref{sec:hydro}), we introduce the hydrodynamic model to
simulate bulk medium expansions.
In Section (\ref{sec:LICM}), the formation of tetraquark and molecular states are
studied with the ICM.
We employ the
test particle Monte Carlo method to numerically solve the combined model
of the Langevin equation and ICM (LICM).
In Section (\ref{sec:physics}), we calculate the $D^0$ spectrum and $J/\psi$ spectrum as a benchmark.
Then, we extend the LICM to the predictions of X(3872) as a compact tetraquark
and a hadronic molecule, respectively.
Their production as a function of the collision centrality and transverse
momentum are analyzed in detail. The final conclusion
is given in Section (\ref{sec:sum}).

\section{Charm quark evolution in phase space}
\label{sec:charm}

\subsection{Langevin equations for charm diffusion}
\label{sec:langevin}

In a hot medium,
with the assumption of
small momentum transfer in each particle scattering,
heavy quark trajectories can
be treated as Brownian motion.
Heavy quark energy loss/gain 
in QGP is attributed to two processes:
elastic scattering
with light partons from the medium and medium-induced gluon radiation.
From previous studies on D meson spectra
from nucleus-nucleus collisions~\cite{Moore:2004tg,
Cao:2013ita,He:2014cla,Lang:2016jpe,Li:2020umn,Li:2021nim},
heavy quark energy loss is dominated by gluon radiation at
high transverse momentum $p_T$~\cite{Guo:2000nz,Zhang:2003wk}
and elastic scattering at low $p_T$~\cite{Qin:2007rn}.
When the local temperatures of QGP drop to the critical temperature $T_c$ of the phase transition,
the heavy quark potential is partially restored. Charmonia can be generated via the coalescence
of $c$ and $\bar c$ quarks in QGP
~\cite{Yao:2020xzw,Yao:2021lus, Zhao:2011cv}.
The charmonia from the coalescence process are mainly located at
low $p_T$ and dominate the final charmonium production.
We focus on the production of charmonia and $X(3872)$ at small and moderate $p_T$
regions, where the gluon radiation effect is negligible. 
The evolutions of charm quarks in QGP can be described by the Langevin equation
~\cite{He:2011qa,He:2013zua,Cao:2015hia},
\begin{align}
{d{\bf p}\over dt}= -\eta {\bf p} + {\bf  \xi}, 
\label{eq-lan}
\end{align}
where $\bf p$ is the heavy quark momentum.
$\eta$ and $\bf \xi$ are the drag and random force terms due to the
interactions with the bulk medium, respectively. In the meantime, the evolution of 
D meson in hadronic medium can also be simulated via Langevin equation. 
The values of the drag force $\eta$ can be obtained through the
fluctuation-dissipation relation,
$\eta = \kappa/( 2T E)$, where
$E=\sqrt{m^2 +|{\bf p}|^2}$ is the heavy quark (or D meson) energy.
$m$ is the mass
of the particle. The charm quark mass is set to $m_c=1.5$ GeV, and
the D meson mass is set to
$m_{D^0}=1.875$ GeV and $m_{\bar D^{*0}}=2.007$ GeV~\cite{Tanabashi:2018oca}.
$T$ is the local temperature of the medium where a heavy quark (D meson) is located.
The momentum diffusion coefficient $\kappa$ is connected with the spatial
diffusion coefficient $\mathcal{D}_s$ through the relation $\mathcal{D}_s\kappa = 2T^2$.
Lattice QCD and effective model calculations indicate that
$\mathcal{D}_s$ is approximately $\mathcal{D}_s(2\pi T)\simeq(4\sim 10)$
at a temperature of approximately $T_c$; see the following review
paper:~\cite{Rapp:2018qla,Dong:2019unq,Zhao:2020jqu}.
Instead of considering the detailed temperature dependence in $\mathcal{D}_s$,
we estimate mean values $\mathcal{D}_s(2\pi T)=5$ for
charm quark diffusion in QGP~\cite{Cao:2015hia}
and $\mathcal{D}_s(2\pi T)=8$ for D meson diffusion in a hadronic medium~\cite{He:2012df}. The $\kappa$ is connected with the random force in Eq. (\ref{eq-lan}) through
\begin{align}  
\langle \xi^{i}(t)\xi^{j}(t^\prime)\rangle =\kappa \delta ^{ij}\delta(t-t^\prime)
\end{align}
with $i,j=(1,2,3)$ representing three dimensions.
$t$ is the time of heavy quark (D meson) evolution in a hot medium.

The Brownian motion of
heavy quarks and D mesons is simulated via
the test particle Monte Carlo method.
At each time step,
Eq. (\ref{eq-lan}) is discretized as
\begin{align}
\label{eq-lan-p}
&{\bf p}(t+\Delta t) ={\bf p}(t)-\eta(p){\bf p}(t) \Delta t+{\bf \xi}\Delta t \\
\label{eq-lan-x}
&{\bf x}(t+\Delta t) = {\bf x}(t) + {{\bf p}(t)\over E}\Delta t \\
&\langle \xi^{i}(t)\xi^j(t-n\Delta t)\rangle ={\kappa \over \Delta t}\delta ^{ij}\delta^{0n} 
\label{fun-dis-lan}
\end{align}
where $n$ is an integer. ${\bf x}(t)$ is the position of the heavy quark at time
$t$. Both the heavy quark momentum and position are updated at each time step with
Eq. (\ref{eq-lan-p}-\ref{eq-lan-x}).
The random noise term in Eq. (\ref{fun-dis-lan}) is set to
a Gaussian distribution with the
width $\sqrt{\kappa/\Delta t}$.

\subsection{Charm quark distributions in Pb-Pb collisions}
\label{sec:charmpt} 

\begin{figure}[!hbt]
\centering
\includegraphics[width=0.48\textwidth]{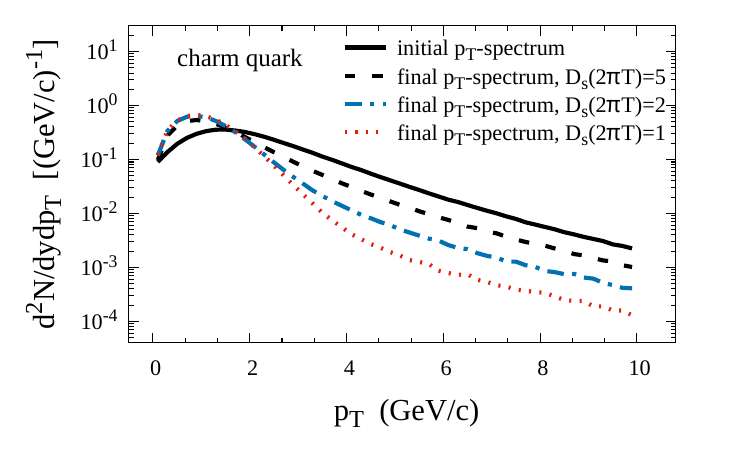}
\caption{The normalized momentum distribution of charm quarks in $\sqrt{s_{NN}}=5.02$ TeV
pp collisions in the central rapidity $|y|<0.9$. The initial momentum
distribution is plotted with a black solid line.
The $p_T$ spectra of charm quarks before the coalescence
are recorded and are plotted with dashed, dotted-dashed and dotted lines.
The different values of the spatial diffusion coefficient correspond to
different degrees of charm kinetic thermalization.
}
\hspace{-0.1mm}
\label{fig-charm}
\end{figure}

The initial momentum distribution of charm quarks in pp collisions at $\sqrt{s_{NN}}=5.02$ TeV is obtained with
the FONLL model{~\cite{FONLL}}.
In Fig. \ref{fig-charm}, the normalized momentum distribution of charm quarks
$d^2N^{\rm norm}_{pp}/dydp_T$ in
the central rapidity $|y|<0.9$
is plotted with a black solid line. The charm quark production cross-section is set to
$d\sigma_{pp}^{c\bar c}/dy=1.165$ mb~\cite{Acharya:2021set}.
In test particle Monte Carlo simulations, the initial momentum of each test particle is
randomly generated based on the black solid line plotted in Fig. \ref{fig-charm}.
Then, the charm quarks exhibit Brownian motion with significant energy loss in the medium.
When the charm quarks move to the positions where the local temperature of the medium is
lower than the hadronization temperature of a certain hadron
such as $T_{c\bar c\rightarrow J/\psi}$, charm and anticharm quarks may combine to form
a new bound state. In Fig.~\ref{fig-charm},
we plot the momentum distributions of charm quarks before and after the evolution 
in QGP by taking different spatial diffusion coefficient 
$\mathcal{D}_s(2\pi T)=1,2,5$. 
With smaller value of
the spatial diffusion coefficient,
charm quarks are closer to the limit of complete momentum thermalization. Heavy quarks
at high $p_T$ are shifted to moderate and low $p_T$ regions.
From the experimental
and theoretical studies of D mesons~\cite{Acharya:2018hre,Cao:2013ita},
the charm quark diffusion coefficient in QGP
is close to $\mathcal{D}_s(2\pi T)=5$.
This diffusion coefficient value is used in the X(3872) yield predictions.

%----------------------------------------
%----------------------------------------

In Pb-Pb collisions,
the initial spatial densities of heavy quarks are
proportional to the number of nuclear binary collisions.
Therefore, the initial positions of the test particles are randomly generated
based on the relative distribution:
\begin{align}
{dN^{\mathrm{test}}\over d{\bf x_T} } \propto T_{A}({\bf x_T}- {{\bf b}\over 2})
T_{B}({\bf x_T} + {{\bf b}\over 2}). 
\label{eq-initx}
\end{align}
where $T_{A(B)}({\bf x_T})=\int dz \rho({\bf  x})$
is the thickness function
of two nuclei and ${\bf x}_T$ is transverse position. The nuclear density
$\rho({\bf x})$ is set to the Wood-Saxon distribution.
${\bf b}$ is the impact parameter, defined as the distance between the centers of two nuclei.
With the momentum and spatial distributions
given in Fig.~\ref{fig-charm} and Eq. (\ref{eq-initx}),
we can randomly generate the initial momentum and initial position
for each test particle and then evolve them event-by-event via
Eq. (\ref{eq-lan-p}-\ref{eq-lan-x}).

\section{Hydrodynamic model for bulk medium evolution}
\label{sec:hydro}

\begin{figure}[!hbt]
\centering
\includegraphics[width=0.49\textwidth]{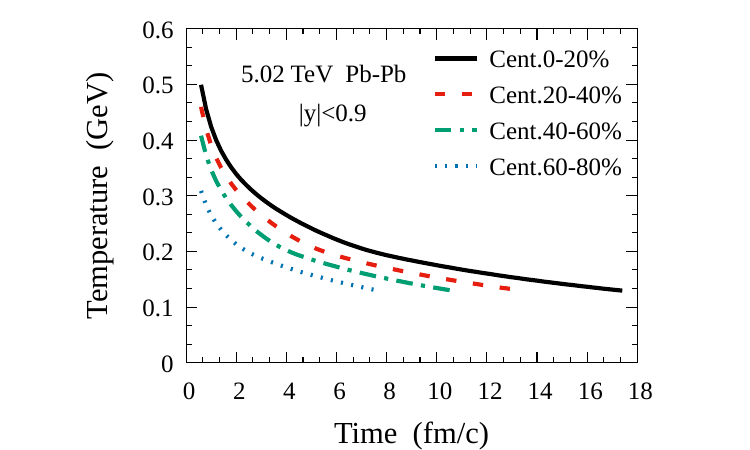}
\caption{Time evolution of the local temperatures at the center of the
hot medium produced in the centralities
0-20\%, 20-40\%, 40-60\%, and 60-80\%.
The hot medium is produced in the central rapidity
in $\sqrt{s_{NN}}=5.02$ TeV Pb-Pb collisions. The results are from the MUSIC
model~\cite{Schenke:2010rr,Schenke:2010nt}.
}
\hspace{-0.1mm}
\label{fig-temp}
\end{figure}

QGP produced in relativistic heavy-ion collisions is a strong coupling medium.
Its expansion can be described with hydrodynamic equations.
In this paper, we employ the (2+1)-dimensional hydrodynamic model to
characterize the time and spatial dependence of the temperatures and
velocities of a hot medium
via the MUSIC package~\cite{Schenke:2010rr,Schenke:2010nt}.
The viscosities of the medium are set to zero for simplicity.
To close the hydrodynamic equations, the equation of state (EoS)
of the medium is needed and can be parametrized via the interpolation between lattice EoS for the deconfined phase and the hadron resonance gas EoS for the hadron phase~\cite{Huovinen:2009yb}.
The two phases are connected with a crossover phase transition.
The hot medium is treated as QGP at $T\ge T_c$ and hadronic
gas at $T<T_c$, respectively.
The critical temperature $T_c$ between QGP and a
hadronic medium is taken to be $T_c=170$ MeV.
The initial maximum temperature at the center of the hot medium is
$T_0(\tau_0, {\bf x_T}=0)=510$ MeV
in the most central collisions ({\bf b}=0)~\cite{Chen:2018kfo}.
The hot medium reaches local equilibrium at $\tau_0=0.6$ fm/c, where
the hydrodynamic
equations start.
In Fig.~\ref{fig-temp},
we plot the time evolution of the QGP
local temperatures at the center of the hot medium at different collision centralities.

\section{The production of heavy flavor hadrons}
\label{sec:LICM}

As QGP expansion, the local temperature will drops continuously. When the temperature
is lower than the dissociation temperature, where the bound state is disappear
due to color screening or scattering, the heavy quark may form a heavy-flavor hadron via
color recombination with other quarks in QGP.
The hadronization process
has been studied with the ICM~\cite{Greco:2003vf,Han:2016uhh,Zhao:2017yan,Zhao:2019ehg} and
resonance recombination model (RRM)~\cite{He:2011qa}. In this section, we
introduce the extended model LICM to study the production
of compact tetraquark and molecular states in heavy-ion collisions.
More discussion on the coalescence model and the statistical model for exotic hadrons
has been presented in a previous review~\cite{Cho:2017dcy}.

\subsection{Charmonium and compact tetraquark states} 

In a medium with a low temperature,
when the relative distance and relative momentum between
charm and anticharm quarks become small,
they may combine into a bound state. The coalescence probability is determined
by the Wigner function of the formed charmonium. We randomly generate $c$
and $\bar c$ quarks based on the distributions in Section (\ref{sec:charmpt});
the ensemble averaged coalescence probability
between uncorrelated $c$ and $\bar c$ in the reaction $c+\bar c\rightarrow \psi +g$
is written as
\begin{align}
\label{eq-psicoal}
&\langle \mathcal{P}_{c\bar c\rightarrow \psi}({\bf x_M}, {\bf p_M})\rangle_{\rm events} 
\nonumber \\
%\langle{d^2P_{c+\bar c\rightarrow \psi}^{\mathrm{test}}\over
%d {\vec x}_M d{\vec p}_M}\rangle_{\rm events} &=
&=g_M \int d{\bf x_1} d{\bf x_2}
{d{\bf p_1}\over (2\pi)^3} {d{\bf p_2}\over 
(2\pi)^3} {d^2N^{\rm test}_1\over d{\bf x_1}d{\bf p_1}} 
{d^2N^{\rm test}_2\over d{\bf x_2} d{\bf p_2}} 
f_M^W({\bf x_r}, {\bf q_r})   \nonumber \\
&\qquad\qquad \qquad \times
\delta^{(3)}({\bf p_M} -{\bf p_1}-{\bf p_2}) 
\delta^{(3)}({\bf x_M} - {{\bf x_1} +{\bf x_2}\over 2}), 
\end{align}
where $g_M=1/12$ for $J/\psi$ is the statistical factor
from color-spin degeneracy.
$d^2N^{\rm test}_i/ d{\bf x_i}d{\bf p_i}$ (i=1,2) are the distributions
of the test particles representing charm and anticharm quarks.
${\bf p_M}$ and ${\bf x_M}$ are the momentum and the coordinate
of the position of the formed charmonium $\psi$.
The momentum carried by the emitting gluon
in the coalescence reaction is neglected, yielding the relation
${\bf p_M} = {\bf p_1}+{\bf p_2}$. The center of the formed charmonium
${\bf x_M}=({\bf x_1}+{\bf x_2})/2$ is located at the middle point
between the charm
and anticharm quarks.
$\langle \mathcal{P}_{c\bar c\rightarrow \psi}({\bf x_M},{\bf p_M})\rangle_{\rm events}$ is the
ensemble averaged coalescence probability between uncorrelated $c$ and $\bar c$ quarks.
The Wigner function
$f_M^W({\bf x_r}, {\bf q_r})$, which serves as the quark coalescence probability, can
be constructed via the Weyl-Wigner transform of the charmonium wave function.
It satisfies the normalization $\int d{\bf x_r}{d{\bf q_r}\over (2\pi)^3} f_M^W({\bf x_r},{\bf q_r})=1$.
For the ground state $J/\psi$, we take the wave function as a 
a simple harmonic oscillator, and the corresponding Wigner function $f^W_M({\bf x_r},{\bf q_r})$
becomes~\cite{Greco:2003vf},
\begin{align}
\label{eq-wig}
f_M^W({\bf x_r}, {\bf q_r})&
= 8\exp[-{{x_r}^{2}\over \sigma^2} - \sigma^2 {q_r}^2]
\end{align}
where the width
can be determined via the mean square radius of the formed
particle
$\sigma^2= {4\over 3}{(m_1+m_2)^2\over m_1^2 +m_2^2}
\langle r^2\rangle_M $~\cite{Greco:2003vf,Song:2016lfv}.
The root-mean-square radius of $J/\psi$ is taken as
$\sqrt{\langle r^2\rangle_{J/\psi}}=0.54$ fm based on the potential model~\cite{Zhao:2020jqu}. 
${\bf x_r}$ and ${\bf q_r}$ are the relative coordinate and relative momentum between two
constituent quarks in the center of mass frame.
Therefore, the coordinates $({\bf x_{1,2}},{\bf p_{1,2}})$ in Eq. (\ref{eq-psicoal})
should be boosted into the center of mass frame (${\bf x}_{1,2}^{cm}, {\bf p}_{1,2}^{cm}$)
before substitution 
into the Wigner function:
\begin{align}
{\bf x_r} &\equiv {\bf x}_1^{cm} - {\bf x}_2^{cm},  \\
{\bf q_r} &\equiv {E_1^{cm}{\bf p}_1^{cm} - E_2^{cm} {\bf p}_2^{cm} \over E_1^{cm} + E_2^{cm}}, 
\label{eq-rela}
\end{align}
where $E_i^{cm}=\sqrt{m_c^2+|{\bf p}_i^{cm}|^2}$ is the energy of the
heavy (anti)quark.
Note that in the event-by-event simulation,
the uncorrelated $c$ and $\bar c$ quarks are unlikely to move
to the QGP fluid cells with the coalescence temperature
$T_{c\bar c\rightarrow \psi}$ at the same time.
The value of the Wigner function decreases rapidly when the relative
distance ${\bf x_r}$ becomes larger than the typical geometry
size of the formed hadron,
which guarantees that in events where hadrons are formed,
$c$ and $\bar c$ quarks are located close to each
other, and their local temperatures are almost the same.
Heavy quarks are rare particles in QGP.
The combination probability of one
$c$ and one $\bar c$ is on the order of magnitude of $\sim 1\%$ in
relativistic heavy-ion collisions. This makes charmonium production
far below the yield of D mesons produced via heavy-light quark coalescence.

With the coalescence probability of random $c$ and $\bar c$ in the hot medium, we can
obtain the $J/\psi$ production in Pb-Pb collisions, which is proportional to the
square of the charm pair numbers:

{
\begin{align}
\label{eq-hadron}
&{ N_{M}^{AA}} = 
\int d{\bf x_M} { d{\bf p_M}\over (2\pi)^3}
\langle{\mathcal{P}_{c\bar c\rightarrow \psi}({\bf x_M},{\bf p_M})}\rangle_{\rm events}
%\nonumber\\&\qquad\qquad\qquad \times 
{( N_{c\bar c}^{AA})^2 } ,\\
& N_{c\bar c}^{AA} = \int d{\bf x_T} T_A({\bf x_T} -{{\bf  b}\over 2})
T_B({\bf x_T} +{{\bf  b}\over 2}) 
%&\qquad \qquad  \times
\mathcal{R}_S 
%\nonumber \\&\qquad\qquad\qquad \times 
{d\sigma_{pp}^{c\bar c}\over dy} \Delta y_{c\bar c},
\end{align}
}
where ${\bf p_M}$ is the three-component momentum
of the formed charmonium respectively.
$ N_{c\bar c}^{AA}$ is the number of charm pairs produced in
the rapidity range $\Delta y_{c\bar c}$ in nucleus-nucleus collisions. As the
coalescence probability decreases with increasing relative rapidity
between charm and anticharm quarks, only those charm and anticharm quarks with
similar rapidities may combine to form a bound state.
$d\sigma_{pp}^{c\bar c}/dy$ is the differential
cross-section of charm pairs in pp collisions.
$ \mathcal{R}_S({\bf b},{\bf x_T})$ is the momentum-averaged nuclear shadowing factor of
charm pairs in the central rapidity of 
Pb-Pb collisions. 
This term is calculated with the EPS09 model~\cite{Eskola:2009uj,Chen:2016dke}. 
The charm pair number is reduced by approximately $\sim 28\%$ after considering
the shadowing effect in the centrality 0-20\%. This
effect becomes weaker in more peripheral collisions.

The above procedure can be extended to four-body coalescence~\cite{Chen:2007zp}. 
We separate the  
judgement of tetraquark coalescence into two parts: 
the relative momentum and coordinate 
between heavy and light quarks need to satisfy the Wigner function, and two 
``diquark''s also satisfy the Wigner function. This coalescence process happens 
instantaneously at $T_{c\bar cq\bar q\rightarrow X}$,
which is the hadronization temperature of the tetraquark.
The light quark position is chosen to be the same as the heavy quark
in the coalescence process, and
its momentum ${\bf p}_{\zeta}^\mathrm{lrf}$ in the local rest frame (LRF) of the
QGP is taken as a normalized
Fermi-distribution ($\zeta$ represents a light quark):
\begin{align}
f({\bf p}_{\zeta}^\mathrm{lrf}) = {N_0 \over e^{\sqrt{m_\zeta^2 + |{\bf p}_{\zeta}^\mathrm{\ lrf}|^2}/T}+1}, 
\label{eq-fermi}
\end{align}
where $T$ is taken to be the coalescence temperature of X(3872).
$N_0$ is the normalization factor. In the event-by-event Monte Carlo simulations,
the light quark momentum in the
LRF of the QGP fluid cell is randomly generated with the normalized
distribution from
Eq. (\ref{eq-fermi}).
It is then boosted to
to the lab frame ${\bf p}_{\zeta}^\mathrm{lab}$. The
diquark momentum is defined as
${\bf p}_\mathrm{diquark}= {\bf p}_c + {\bf p}_{\zeta}^\mathrm{lab}$.
Note that the blueshift
of the light quark momentum distribution due to the moving fluid cells
is consistently included by boosting the light quark momentum
from ${\bf p}_{\zeta}^\mathrm{lrf}$
to ${\bf p}_{\zeta}^\mathrm{lab}$.
The four velocities
of the QGP fluid cells and the QGP temperature profile $T({\bf x_T}, t)$ are
given by the hydrodynamic model.
The light quark thermal mass in the hot medium
is set to $m_\zeta=0.3$ GeV in Eq. (\ref{eq-fermi}).

With the momenta and coordinates of the diquark and antidiquark,
the formation of a tetraquark state is similar to the $J/\psi$ situation.
With X(3872) spin of
$J_{X(3872)}=1$, the tetraquark statistical factor from color-spin degeneracy is
$g_{X(3872)}=1/432$ in the coalescence equation.
The form of the X(3872) Wigner function is that of Eq. (\ref{eq-wig}).
With the assumption that the tetraquark root-mean-square radius 
is approximately $0.3\sim 0.5$ fm which is similar to $J/\psi$,
we take the coalescence
temperature of the tetraquark X(3872) to be
$T_{ c\bar cq\bar q\rightarrow X}\simeq T_{c\bar c\rightarrow J/\psi}
\simeq 1.2T_c$, slightly above the critical
temperature $T_c$~\cite{Satz:2005hx,Guo:2019twa,Zhao:2020jqu}.

\subsection{Molecular states}

\begin{figure}[!hbt]
\centering
\includegraphics[width=0.49\textwidth]{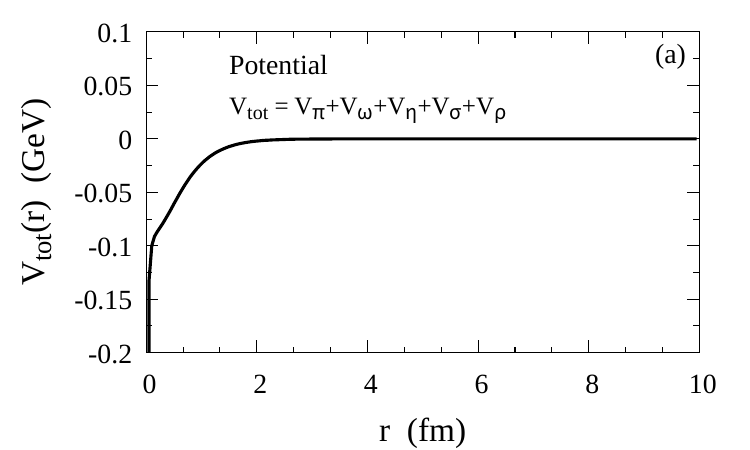}
\hfill
\includegraphics[width=0.49\textwidth]{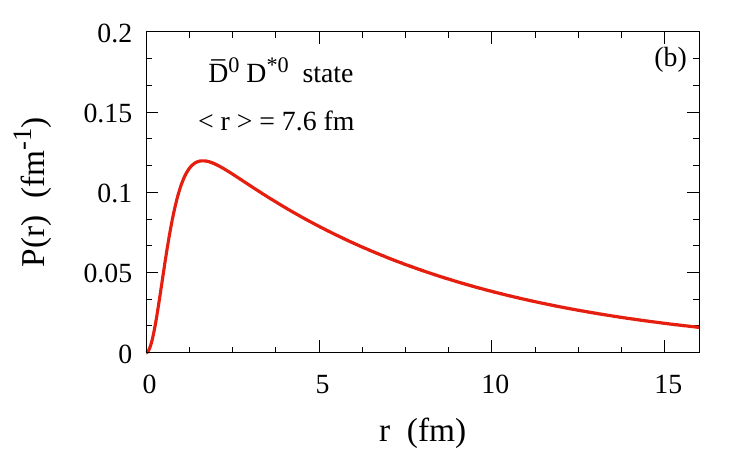}
\caption{Upper panel (a): Total potential between $\bar D^0$ and $D^{*0}$
(or $D^0$ and $\bar D^{*0}$) in the molecular state.
Lower panel (b): the radial probability
$P(r)=|\psi(r)|^2r^2$ of the loosely bound molecular X(3872).
}
\hspace{-0.1mm}
\label{fig-DD}
\end{figure}

As the mass of X(3872) is very close to the threshold
mass of $D^0\bar D^{*0}$ (or $\bar D^0 D^{*0}$), it has been
suggested that X(3872) is a loosely bound molecular state~\cite{Liu:2008tn}.
The interaction potential between different D mesons can be obtained
with the effective Lagrangians, including the
contributions from the exchanges of $\pi, \eta, \sigma, \rho$ and $\omega$ mesons.
The total effective potential for $D^0 \bar D^{*0}$ (or $\bar D^0 D^{*0}$) is attractive,
as shown in~\cite{Liu:2008tn}. To regularize the potential,
we impose a short-distance cutoff $\Lambda$ to address the singularity
of the effective potential. The cutoff $\Lambda$ affects the range
of the interaction potential. Solving the two-body Schr\"odinger equation with this potential,
we obtain the wave function and the binding energy of the
$D^0 \bar D^{*0}$ molecular state, as shown in table~\ref{tab-mole} and Fig.~\ref{fig-DD}.
The definition of the binding energy is the mass difference
between X(3872) and $\bar D^0 D^{*0}$; i.e.
$\text{BE.}\equiv  M_{\bar D^0}+M_{D^{*0}}-M_{X(3872)}$.
%---------------------------------------------------------------------
\begin{table*}[!hbt]
\renewcommand\arraystretch{1.5}
\centering
\setlength{\tabcolsep}{2.5mm}
\begin{tabular}{c|ccccccc}
	\toprule[1pt]\toprule[1pt] 
	 $\Lambda$& 0.55  &0.555 & 0.56 & 0.565 & 0.57 & 0.575 & 0.579 \tabularnewline
	\midrule[0.6pt]
	\text{BE.}(keV) & 1600.3 & 1098.5& 698.4 & 394.4& 180.6& 51.2 & 3.3  \tabularnewline
	$\langle r \rangle$(fm) & 2.47 & 2.85 & 3.41& 4.31& 6.01 & 10.52 & 22.60  \tabularnewline
	$\sqrt{\langle r^2 \rangle}$($\mathrm{fm}$) & 3.08 & 3.59 & 4.36& 5.61 & 8.00 & 14.33 & 28.94  \tabularnewline
	\bottomrule[1pt]\bottomrule[1pt]
	\end{tabular}
\caption{The binding energy and the mean radius (and root-mean-square radius) 
of molecular state X(3872) 
with different values of the parameter $\Lambda$ in the potential.}
\label{tab-mole}
\end{table*}
%---------------------------------------------------------------------
%--

From the table, we can see that the binding energy
and the mean radius are sensitive to the interaction potential
and the cutoff parameter $\Lambda$.
%The binding energy of X(3872)
%is roughly in the range of 10-1000 keV~\cite{Cho:2017dcy,Tanabashi:2018oca}.
In Fig. \ref{fig-DD}, the binding energy is set to 100 keV,
the corresponding potential and the wave function are plotted with the
mean radius of $\sim 7.6$ fm.

%----------------------------------------

In heavy-ion collisions with the production of a hot medium,
(anti)charm quarks first form $D$(or $\bar D$) mesons
at the QCD phase boundary. Then, in the hadronic phase,
the D mesons continue diffusing.
Due to the low binding energy
of the molecular state, molecular X(3872) can be formed only
via the coalescence of
$D^0$ and $\bar D^{*0}$ mesons after the medium reaches
kinetic freeze-out.
No molecular states can survive in the hadronic medium
above the temperature $T_{\rm kin}$ of the kinetic freeze-out due to the
random scattering with surrounding light hadrons.
In this section, we extend the coalescence
model to the molecular formation.
In the low and moderate $p_T$ regions, charm and light quarks can form D mesons
at the critical temperature $T_c$.
The coalescence formula for D meson is written as (taking $D^0$ as an example),
{\small
\begin{align}
\label{eq-Dcoal}
&\langle\mathcal{P}_{c\bar q\rightarrow D^0}({\bf p_D})\rangle_{\rm events}\nonumber \\
&= \mathcal{H}_{c\rightarrow D^0} 
\int {d{\bf p_1}\over (2\pi)^3} {d{\bf p_2}\over (2\pi)^3} 
%\nonumber \\&\times
{dN_1\over d{\bf p_1}} {dN_2\over d{\bf p_2} }
f_D^W({\bf q_r})\delta^{(3)}({\bf p_D} -{\bf p_1}-{\bf p_2}),  \\
\label{eq-DAA}
& N_{D^0}  = 
\int { d{\bf p_D}\over (2\pi)^3}
\langle{\mathcal{P}_{c\bar q\rightarrow D^0}({\bf p_D})}\rangle_{\rm events} 
%\times 
{N_{c\bar c}^{AA} },
\end{align}}
\noindent 
where $\mathcal{H}_{c\rightarrow D^0}$ is the hadronization ratio of charm quarks
turning into a direct $D^0$ state (similar for $D^+$, $D^{*0}$, $\Lambda_c$, etc.) We take the
values of the hadronization ratios to be $\mathcal{H}_{c\rightarrow D^0}=9.5\%$ and
$\mathcal{H}_{c\rightarrow D^{*0}}=20\%$~\cite{Acharya:2018hre}.
The same hadronization ratios
are used in the coalescence of $\bar D^0$ and $\bar D^{*0}$ mesons.
${\bf p_D}$ is the momentum of the formed D meson. 
$\langle\mathcal{P}_{c\bar q\rightarrow D^0}({\bf p_D})\rangle_{\rm events}$ is the
ensemble-averaged probability of charm quarks turning into $D^0$ mesons. We assume
that all D mesons are produced via the coalescence process and neglect the
fragmentation contribution.
The integration of Eq. (\ref{eq-Dcoal}) over momentum is 1.
$dN_i/d{\bf p}_i$ (i=1,2) represents the momentum
distributions of two test particles at the positions with the
coalescence temperature of D mesons $T=T_c$.
The position and momentum of charm quarks at the hadronization surface
can be obtained via the Langevin equations given in Eqs. (\ref{eq-lan-p}-\ref{fun-dis-lan}).
The momentum distribution of the light quark in the LRF of the
QGP in Eq. (\ref{eq-Dcoal}) is taken to be the normalized
Fermi-distribution; see Eq. (\ref{eq-fermi}).
The width of the D meson Wigner function is
determined in the same way as that of charmonium, and the
root-mean-square radius of the D mesons
is taken as $\sqrt{\langle r^2\rangle_D}=0.43\ \mathrm{fm}$~\cite{Zhao:2020jqu} for both
$D^0$ and $D^{*0}$.

After the formation of D mesons, they continue to diffuse in the hadronic medium,
with a different value for the spatial diffusion coefficient, $\mathcal{D}_s(2\pi T)=8$, 
as discussed in Section~\ref{sec:charm}.
When D mesons move to the regions where the hadronic medium reaches
kinetic freeze-out, $D^0$ and $\bar D^{*0}$ mesons may combine to form a
loosely bound molecular state.
We set the coalescence temperature of molecular X(3872) to be the
kinetic freeze-out temperature, $T_{\rm mole}\simeq T_{\rm kin}\simeq 0.14$ GeV
(the kinetic freeze-out temperature can be extracted from experimental
data~\cite{Adamczyk:2017iwn}).
Due to the uncertainty of
the molecular geometric size
given in table~\ref{tab-mole}, we take its value to
be  $\sqrt{\langle r^2\rangle_{X}}=3.0, 5.5, 9.0$ fm in the calculations of hadronic 
molecule production.
Actually, the strategy we used here for the molecular state of X(3872) is similar to the light nuclei production in heavy-ion collisions~\cite{Zhao:2021dka}. The proton and neutron are formed in the QCD phase boundary (also many feed-down contributions in the hadronic phase) and evolve in the hadronic phase. When the system undergoes the chemical freeze-out, the coalescence of light nuclei, such as deuteron and triton, happens.

%{\color{red}{Other composite systems like Helium3 or Hypertriton can also be formed in the heavy 
%ion collisions~\cite{Esposito:2015fsa}. The coalescence processes of Helium3 and the 
%molecular 
%X(3872) are similar to each other. Light quarks combine to form a proton or a 
%neutron at the hadronization surface, and prtons and neutrons combine to form Helium3 or 
%Hypertriton in the hadronic medium. 
%This is similar to the formation process of the molecular X(3872). 
%One of the main difference is their coalescence ingredient.  
%In the QGP, light quarks can be easily thermally produced as their masses are smaller 
%than the medium temperature. This significantly enhances the final production 
%of Helium3 and Hypertriton in Pb-Pb collisions compared with the situation of pp collisions. 
%X(3872) consists of charm and anti-charm quarks, which 
%are hardly produced via random collisions of thermal partons in QGP. The number 
%of charm pairs in the QGP is nearly conserved. The enhancement of X(3872) yield is not 
%so significant as the light composite systems. }} 
%

\subsection{Numerical simulations}

We employ the test particle Monte Carlo method to numerically solve
the LICM.
In each event, two test particles are randomly generated with
uncorrelated initial positions and initial momenta. Their dynamical evolution is
described with two independent 
Langevin equations. When they move to the regions
where local medium temperatures drop to the coalescence temperature,
their relative distance and relative momentum are calculated; these
parameters are
used in the Wigner functions to calculate the probability of
coalescence that forms a
new bound state.
With the coalescence probability between two test particles, we
generate a random number between 0 and 1 and compare it with the coalescence
probability. If the coalescence probability is larger
than this random number, the new bound state can be formed. Otherwise,
the test particles
continue independently evolving.
In event-by-event simulations, the particle distributions
in Eqs. (\ref{eq-psicoal} and \ref{eq-Dcoal}) become delta functions.
For example,
the charm quark distribution before the coalescence process can be written as
$d^2N_1/d{\bf x_1}d{\bf p_1}=(2\pi)^3
\delta^{(3)}({\bf x_1}-{\bf x_c})\delta^{(3)}({\bf p_1-\bf p_c})$, where
$({\bf x_c},{\bf p_c})$ includes the coordinate and momentum of the charm quark at the moment
of coalescence.

\section{ $D$ meson, charmonium and X(3872) observables}
\label{sec:physics}
In the above sections, we introduced the LICM to describe the diffusions of
charm quarks and D mesons in the hot medium and the coalescence process.
Now, we calculate the spectra of prompt $D^0$ and $J/\psi$ mesons in Pb-Pb collisions
as a benchmark of X(3872) production.
Due to different binding
energies of $D$ and $J/\psi$, they are decoupled with hot medium by different temperatures.

\begin{figure}[!hbt]
\centering
\includegraphics[width=0.44\textwidth]{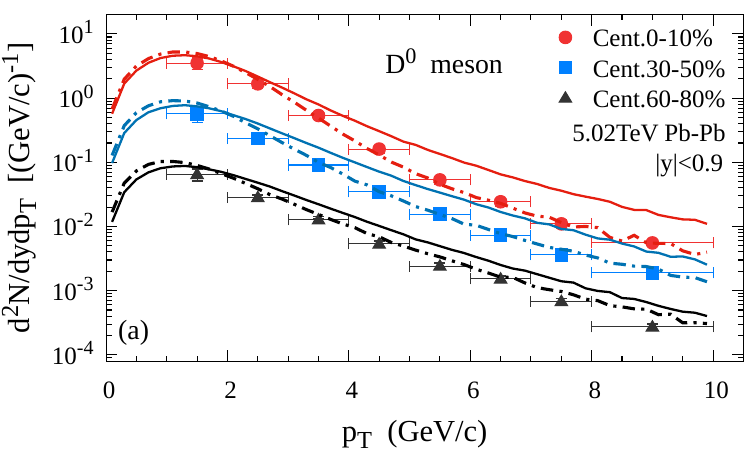}
\hfill
\includegraphics[width=0.44\textwidth]{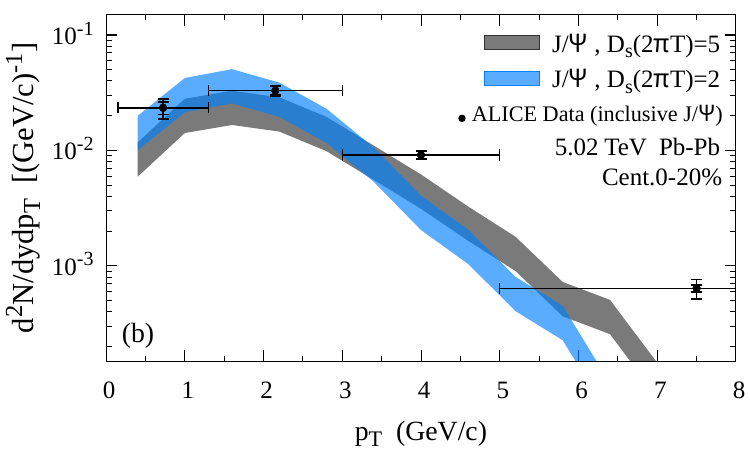}
\caption{ Upper panel (a): Transverse momentum spectra of prompt $D^0$ mesons in different
centralities 
in $\sqrt{s_{NN}}=5.02$ TeV Pb-Pb collisions are plotted. 
The solid line and
dotted-dashed line represent the conditions of $\mathcal{D}_s(2\pi T)=5$ and
$\mathcal{D}_s(2\pi T)=2$, respectively. The cold nuclear matter effects
are included in all the lines.
Lower panel (b):
$J/\psi$ transverse momentum spectra $d^2N/dydp_T$ in Pb-Pb collisions
at $\sqrt{s_{NN}}=5.02$ TeV. The collision centrality is 0-20\%.
The experimental data are the $J/\psi$ inclusive production from the ALICE
Collaboration~\cite{Acharya:2019lkh}. 
The theoretical results are the regeneration production from the coalescence.
The two bands correspond to different values of
charm quark spatial diffusion coefficient in QGP.
The lower and upper limits
of the theoretical bands correspond to situations with and without the shadowing effect,
respectively.
}
\hspace{-0.1mm}
\label{fig-Jpsipt}
\end{figure}

In the prompt $D^0$ spectrum in Fig.\ref{fig-Jpsipt}, as we do not include radiative energy
loss, the theoretical
calculations with $\mathcal{D}_s(2\pi T)=5$ (solid lines) underestimate the energy loss
of $D^0$ mesons at high $p_T$. As a compensation,
$\mathcal{D}_s(2\pi T)=2$ (dotted-dashed lines)
is also taken. This can significantly change the spectrum of
$D^0$ mesons at high $p_T$ but becomes negligible at low $p_T$, as we expected.
The ratio of prompt $D^0$ meson over total charm number is determined with the
ratio given in pp collisions $N_{D^0}^{\rm prompt}/N_{c\bar c}=39\%$~\cite{Acharya:2021set}. 
We focus on the $p_T$-integrated yields of $J/\psi$ and X(3872), which are dominated by the
coalescence process at low and moderate $p_T$. 
For $J/\psi$ experimental data at high $p_T$, the inclusive production is dominated by
the primordial production and B-decay contributions, which are absent in the theoretical
calculations (color bands)~\cite{Chen:2013wmr}. This explains why our $J/\psi$
calculations are lower than the
experimental data at $p_T\gtrsim 4$ GeV/c.
At $p_T\lesssim 4$ GeV/c, our theoretical calculations explain
the experimental data well for both prompt $D^0$ and $J/\psi$.
The lower and upper limits of the color bands in the $J/\psi$ calculations
correspond to the situations with and without the nuclear shadowing effect.

\begin{figure}[!hbt]
\centering
\includegraphics[width=0.44\textwidth]{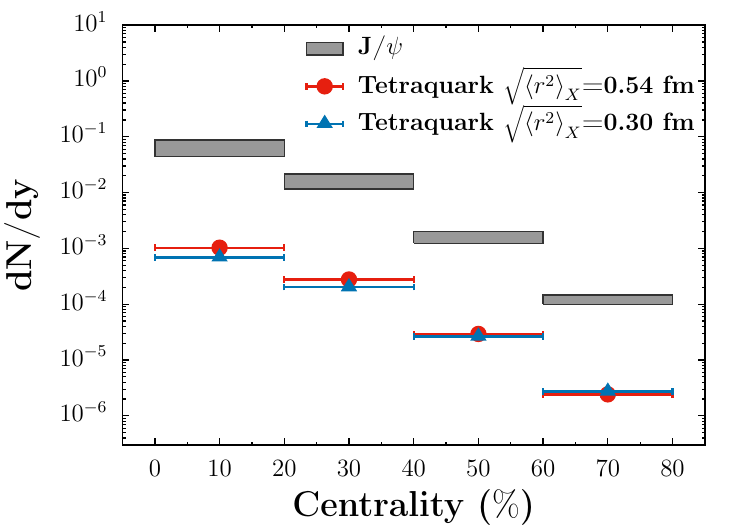}
\caption{
Tetraquark and $J/\psi$ production as a function of the collision centrality
in the central rapidity of Pb-Pb collisions at $\sqrt{s_{NN}}=5.02$ TeV. 
Four centralities are chosen, 0-20\%, 20-40\%, 40-60\%, and 60-80\%.
The band of $J/\psi$ calculations represents the situations with and without the
nuclear shadowing effect. All the tetraquark lines
include the nuclear shadowing effect.
The spatial diffusion coefficient of charm quarks in QGP is set
to $\mathcal{D}_s(2\pi T)=5$.
}
\hspace{-0.1mm}
\label{fig-Np}
\end{figure}

In the formation of a tetraquark, first, a charm quark
combines with a light
quark to form a diquark, and then the diquark and an antidiquark
combine to form a tetraquark at
the coalescence temperature $T_{c\bar cq\bar q\rightarrow X}$.
As light quarks are abundant in QGP,
tetraquark production is mainly determined by the density of charm pairs
and the Wigner function of the tetraquark state.
Different from pp collisions, most 
primordially produced
tetraquarks are melted in QGP due to the strong color screening effect.
The final production of tetraquarks mainly comes from the
coalescence process. We plot tetraquark production as a function of centrality
in Fig. \ref{fig-Np}, and $J/\psi$ production is plotted as a comparison.
The band of $J/\psi$ production represents the situations with and without
the nuclear shadowing effect.
In tetraquark production, different values of the width in the Wigner function
are considered by setting the root-mean-square radius of the tetraquark to
$\sqrt{\langle r^2\rangle_{X}}=0.3$ fm and 0.54 fm (the latter is the same as $J/\psi$).
First, we can see that the $J/\psi$ production is much larger than the tetraquark production. 
This is mainly induced by the different statistical factors in the coalescence equation. 
Our predictions for the tetraquark yield are consistent with Ref.\cite{Cho:2017dcy}.
In Fig. \ref{fig-Np}, when the geometric size of the tetraquark is increased,
its production increases by approximately 40\% in the central collisions.
However, in peripheral collisions, due to the smaller volume and shorter lifetime
of the QGP, charm quarks experience less energy loss in the medium, which increases
the relative momentum between uncorrelated $c$ and $\bar c$.
Considering the relative momentum part of the
Wigner function given in Eq. (\ref{eq-wig}), with a larger mean radius, the tetraquark yield
is more reduced for centrality 60-80\%,
as shown in Fig.~\ref{fig-Np}.

\begin{figure}[!hbt]
\centering
\includegraphics[width=0.44\textwidth]{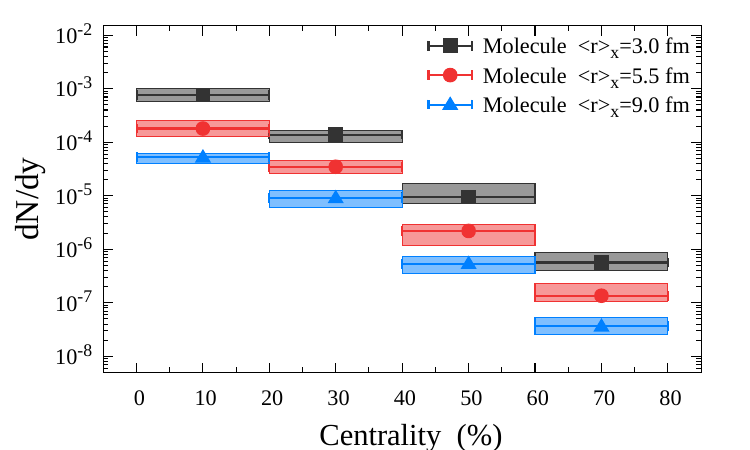}
\caption{ Molecular production as a function of centrality in the central rapidity
in $\sqrt{s_{NN}}=5.02$ TeV Pb-Pb collisions.
Different geometric sizes of the molecular state are considered. The spatial diffusion
coefficients of charm quarks in QGP and D mesons in a hadronic medium are taken
as $\mathcal{D}_s(2\pi T)=5$ and 8, respectively. The upper and lower limits of 
the bands correspond to the different values of the kinetic freeze-out temperature 
$T_{\rm mole}=0.16$ GeV and 0.10 GeV respectively. 
}
\hspace{-0.1mm}
\label{fig-Npmole}
\end{figure}

If X(3872) is a molecular state, its binding energy is on the order of
$\sim$ keV.
X(3872) is then produced
via the coalescence of $D^0$-$\bar D^{*0}$ or $\bar D^0$-$D^{*0}$
in the hadronic medium at the temperature at which the medium reaches kinetic freeze-out
$T_{\rm mole}=0.14$ GeV.
The molecular geometric size is much larger than that of the compact tetraquark. Its mean radius
and the binding energy are calculated based on the potential model in
table~\ref{tab-mole}. Determining the exact value of the X(3872) geometric size is beyond the scope
of this work. Instead, we take different geometric sizes for the hadronic 
molecule and calculate
the X(3872) production.
The root-mean-square radius of the molecular state is set
to $\sqrt{\langle r^2\rangle_{X}}=3.0, 5.5, 9.0$ fm. In Fig.~\ref{fig-Npmole}, the molecular production with
$\sqrt{\langle r^2\rangle_{X}} =3.0$ fm is at the same order as
the tetraquark production.
When the molecular geometric size increases, $\sigma$
in the Wigner function also increases. 
This gives strict momentum conditions in the coalescence of $D^0$ and $\bar D^{*0}$ mesons.
Only when $D^0$ and $\bar D^{*0}$ mesons
are separated by a large distance but also carry almost the same
momentum can they
form a molecular state. This constraint
significantly suppresses the molecular yield.
In the limit of the molecular binding energy approaching zero,
the mean radius of the loosely
bound hadronic molecule goes to infinity. This means that $D^0$ and $\bar D^{*0}$
mesons must carry almost
the same momentum to form a molecular state, which makes
the coalescence probability between D mesons very small in a hadronic medium.

Both the tetraquark
and molecular yields in Fig. \ref{fig-Np}-\ref{fig-Npmole} show clear centrality dependence.
They are proportional to the square
of the heavy flavor densities in the hot medium.
In more central collisions, more charm pairs and X(3872) are produced.
This centrality dependence of X(3872) production
is qualitatively consistent with
the rate equation model~\cite{Wu:2020zbx}.
If the relative momentum between D mesons in the center of mass frame of the 
hadronic molecule 
is a few pion mass~\cite{Artoisenet:2009wk}, the 
the molecular root-mean-square radius is taken as $\sim 3$ fm. Molecular
production will be strongly enhanced and become comparable with 
the tetraquark production.
If the molecular geometric size
is larger, the molecular production is several times lower;
see Fig.~\ref{fig-Np}-\ref{fig-Npmole}.
One of the main reasons for this is the Wigner functions used for X(3872) production.
In this model, the molecular formation conditions in both physical and momentum
space are closely connected via one parameter: the width in the Wigner function.
With a very large geometric size for molecular X(3872),
more D and $\bar D$ mesons satisfy the spatial formation conditions, but this also
results in a strict momentum constraint on the momentum part of the Wigner function.
The value of the momentum part of the Wigner function
$\exp(-\sigma^2 q_r^2)$
is significantly reduced when the relative momentum $q_r$ between
$D^0$ and $\bar D^{*0}$ mesons increases.
The consistent constraints from both spatial and momentum formation
conditions result in molecular production not increasing with
geometric size. The freeze-out temperature of the molecular state is a parameter 
in this work and depends on the collision centrality~\cite{ALICE:2019hno}. We assume that 
the values of the 
freeze-out temperature change between 0.16 GeV and 0.10 GeV in different 
collision centralities. The molecular production with different kinetic freeze-out 
temperature is plotted with 
bands in Fig.\ref{fig-Npmole}. With higher freeze-out temperature, molecular production 
is enhanced due to the larger spatial density of the D mesons in the hot medium.

With the production of tetraquark and molecular states in 
Fig.\ref{fig-Np}-\ref{fig-Npmole}, we can obtain  
the nuclear modification factor $R_{AA}^{X(3872)}$ of $X(3872)$.  
First, 
we calculate the nuclear modification factor of $J/\psi$ at 5.02 TeV Pb-Pb collisions. 
Take the differential cross section of 
prompt $J/\psi$ to be $d\sigma_{pp}^{J/\psi}/dy=5.0$ $\mu b$ in the central rapidity,  
$J/\psi$ nuclear modification factor is $0.42<R_{AA}^{J/\psi}<0.81$ in 
the centrality 0-20\%, 
where lower and upper limits correspond to the situations with and without cold nuclear 
matter effect in Fig.\ref{fig-Np}. 
This result is consistent with other theoretical 
calculations~\cite{Zhao:2011cv, Chen:2018kfo}
and the experimental data~\cite{ALICE:2016flj}. 
For the production cross section of X(3872) in pp collisions, the yield ratio 
$N^{X(3872)}_{pp}/N^{\psi(2S)}_{pp}$ has been measured by LHCb Collaboration at 
$\sqrt{s_{NN}}=8$ TeV. 
The central value of the ratio is  
$N^{X(3872)}_{pp}/N^{\psi(2S)}_{pp}\approx 0.1$ at the low multiplicity pp 
collisions~\cite{LHCb:2020sey}. 
We take the same value of 
$N^{X(3872)}_{pp}/N^{\psi(2S)}_{pp}$ at $\sqrt{s_{NN}}=5.02$ TeV, and extract the 
prompt yield ratio 
to be $N^{X(3872)}_{pp}/N^{J/\psi}_{pp}\simeq 8.0\times 10^{-3}$. 
If X(3872) is a tightly bound hadron state, the 
prompt nuclear modification factor satisfies the relation 
$R_{AA}^{X(3872)}/R_{AA}^{J/\psi}\approx 2.8$ in the 
centrality 0-20\% in Pb-Pb collisions, which indicates 
$R_{AA}^{X(3872)}$ to be 
$1.2\sim 2.3$ in the scenario of $\sqrt{\langle r^2\rangle_X}=0.54$ fm. 
If X(3872) is a loosely bound molecular state, the value of 
$R_{AA}^{X(3872)}$ is $0.24\sim 0.46$ in the 
scenario of $\sqrt{\langle r^2\rangle_X}=5.5$ fm.

The ratio between X(3872) and $\psi(2S)$ production in Pb-Pb collisions has also been measured 
by CMS Collaboration~\cite{CMS:2021znk}. 
$\psi(2S)$ prompt production can be estimated 
via a simple thermal weight factor 
$(m_{\psi(2S)}/m_{J/\psi})^{3/2}\exp(-(m_{\psi(2S)}-m_{J/\psi})/T)$~\cite{Greco:2003vf}. 
The temperature 
in the exponential factor is taken as the $J/\psi$ coalescence temperature. 
The yield ratio of $\psi(2S)$ to $J/\psi$ 
is $\sim 7.3\%$. Then we obtain the value of the ratio to be around 
$N_{AA}^{X(3872)}/N_{AA}^{\psi(2S)}\simeq 0.30$ (tetraquark scenario 
with $\sqrt{\langle r^2\rangle_X}=0.54$ fm) and $0.06$ 
(hadronic molecule scenario with $\sqrt{\langle r^2\rangle_X}=5.5$ fm), respectively. 
If the geometry size of the molecular state becomes smaller by taking 
$\sqrt{\langle r^2\rangle_X}$ to be or smaller than 3.0 fm, 
the yield of the molecular state can  become 
larger than 
the tetraquark production. The final production of X(3872) depends on its 
wave function which is characterized by the parameter $\sqrt{\langle r^2\rangle_X}$. 
Note that the yield ratio from above theoretical calculations are in the low $p_T$ region 
where X(3872) and $\psi(2S)$ are mainly from the coalescence process, 
while the experimental 
data in Ref.\cite{CMS:2021znk} are located in high $p_T$ region where X(3872) are 
produced by the primordial 
parton hard scatterings.

\begin{figure}[!hbt]
\centering
\includegraphics[width=0.49\textwidth]{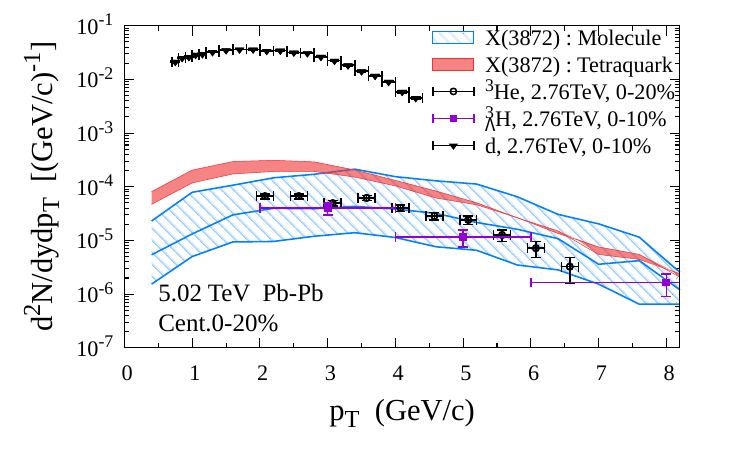}
\caption{ $p_T$ spectra of $X(3872)$ as a tetraquark and a
hadronic molecule
in $\sqrt{s_{NN}}=5.02$ TeV Pb-Pb
collisions. The collision centrality is 0-20\%. The parameters are the
same as those in
Fig.~\ref{fig-Np}-\ref{fig-Npmole}.
The uncertainties of the tetraquark and hadronic 
molecule yields are induced by the different values
of the width in the Wigner function. The upper and lower limits in the tetraquark spectrum
correspond to the root-mean-square values of the tetraquark state of
$\sqrt{\langle r^2\rangle_X}=0.3$ and 0.54 fm (the same as
$J/\psi$). The three lines in
the molecular spectrum correspond to the situations of $\sqrt{\langle r^2\rangle_X}=3.0,
5.5, 9.0$ fm.
The nuclear shadowing effect is included in all the calculations. 
The circular, square and triangle points 
represent the experimental results of $^3$He, $^3_\Lambda$H and $d$ 
in 2.76 TeV Pb-Pb collisions~\cite{ALICE:2015wav,ALICE:2015oer}, respectively. 
}
\hspace{-0.1mm}
\label{fig-X3872-pt}
\end{figure}

In Fig. \ref{fig-X3872-pt}, the $p_T$ spectra of X(3872) as a tetraquark and hadronic 
molecule
are plotted. The uncertainty bands in the theoretical calculations are due to the
different choices for the width in the X(3872) Wigner functions.
With an increasing value for the
width, the tetraquark and hadronic molecule production values show different changes.
Tetraquark production is enhanced, but hadronic molecule production is reduced.
This is due to the
combined effects from the spatial and momentum formation
conditions that are
consistently given via the Wigner function in Eq. (\ref{eq-wig}).
The peak of the molecular $p_T$ spectrum is shifted to larger
$p_T$ compared with that of the tetraquark spectrum.
This is because the molecular state is produced in the later stage of the
hot medium expansion and the $p_T$ of the hadronic 
molecule can be shifted by the radial flows of
the expanding hot medium. With the violent expansion of the hot medium, its radial flows 
increase with time, which will be picked up by charm quarks and D mesons 
via random scatterings with the medium. 
The $p_T$ spectra of different particles produeced at different stages of hot medium expansion 
will be sequentially modified~\cite{ref:sequen}. At very high $p_T$, inspired by 
$J/\psi$ studies, the production of exotic 
heavy flavor hadrons (or hadronic molecules) in the coalescence process is 
believed to become negligible 
compared with the primordial production.  

{{
It is interesting to compare the production of X(3872) with other hadronic molecules (light nuclei), such as deuteron (d), helium-3 ($^3$He), and hypertriton ($^3_\Lambda$H). In the high multiplicity p-p collisions, the comparison has been made, and the results indicate any loosely bound hadronic molecule interpretation of X(3872) is questionable~\cite{Esposito:2015fsa}. Here, we focus on the Pb-Pb collisions. Due to the lack of experimental data in 5.02 TeV Pb-Pb collisions, we add the results of $d$, $^3$He, and $^3_\Lambda$H in 2.76TeV Pb-Pb collisions in Fig.~\ref{fig-X3872-pt}. We can see the yield of $d$ is about 2 orders of magnitude larger to the X(3872) production. 
And the yields of $^3$He and $^3_\Lambda$H are comparable with the molecular-like X(3872) in heavy-ion collisions. 
Even the production mechanism of light nuclei and molecular-like X(3872) are similar in relativistic heavy-ion collision at low $p_T$ region. But the abound of protons and neutrons in the hadronic phase enhance the yield of two-body molecular state, $d$. For the three-body molecular state $^3$He and $^3_\Lambda$H, the coalescence probability constrains the phase-space distribution of protons, neutrons, and hyperons, which in turn reduce their production.
Due to the coalescence production, this behavior is much different from the case in pp collisions, especially in the high $p_T$ region~\cite{Esposito:2015fsa}. Our results show, in relativistic heavy-ion collision, the hadronic molecule interpretation of X(3872) is not excluded in the low $p_T$ region so far.}}

We also check the sensitivity of X(3872) production with the
different choices of parameters. When the coalescence temperature of
the tetraquark state
is shifted to the critical temperature $T_c$,
heavy quarks diffuse to a larger volume in QGP
before forming a tetraquark state. The tetraquark yield is
fractionally suppressed 
due to the smaller spatial density of heavy quarks in the medium.
This effect is similar in the molecular scenario.
Different degrees of heavy quark
kinetic thermalization can also affect the final production values of tetraquarks and hadronic molecules.
In the limit of charm quark kinetic thermalization, both tetraquark and molecular
production can be enhanced by approximately $\sim 2$ times compared with the situations 
in Fig.~\ref{fig-Np}-\ref{fig-Npmole}.
Different from D mesons, the production of X(3872) depends on the
square of the charm pair number in heavy-ion collisions.
The uncertainty in the charm pair production cross-section $d\sigma_{pp}^{c\bar c}/dy$
is amplified in X(3872) production.
The scope of the work is to distinguish the nature of X(3872)
via the geometric size of its wave function, which is one of the most important
differences between the compact tetraquark and the loosely bound hadronic molecule.

\section{Summary}
\label{sec:sum}
In this work, we develop the Langevin equation and instant coalescence model (LICM)
to study the production of open and hidden charm flavors including prompt $D^0$, $J/\psi$
and X(3872) in heavy-ion collisions. Calculations regarding $J/\psi$ and $D^0$ mesons are the
benchmark of our predictions regarding X(3872) as a tetraquark state and a hadronic molecule, respectively.
The realistic diffusions of charm quarks in quark-gluon plasma (QGP) and D mesons in the
hadronic medium are described with the Langevin equation.
The spatial and momentum formation conditions of X(3872)
are consistently given in the Wigner function,
which encodes the internal structure of the formed particle.
The compact tetraquark and loosely bound hadronic molecule
are produced at different medium temperatures: a tetraquark is formed in QGP above the critical
temperature, while a hadronic molecule is formed only in the
hadronic medium after the kinetic freeze-out.
With the constraints of color-spin degeneracy, X(3872) production as a tetraquark
state becomes much smaller than $J/\psi$ production.
The geometric size of molecular state is very large, and its binding energy is almost
zero. This requires the relative momentum between $D^0$ and $\bar D^{*0}$ mesons
to be small to form a loosely bound hadronic molecule.
Strict constraints on the relative momentum in the Wigner function
significantly suppress the molecular yields. Nuclear modification 
factor $R_{AA}^{X(3872)}$ of X(3872) as a tetraquark and molecular states 
are also calculated. Its value becomes $R_{AA}^{X(3872)}>1$ and $<1$ respectively in 
the scenarios of tightly bound state and weakly bound state, which is characterized by 
the parameter of the root-mean-square $\sqrt{\langle r^2\rangle_X}$. 
The ratio $N_{AA}^{X(3872)}/N_{AA}^{\psi(2S)}$ 
between X(3872) and $\psi(2S)$ production in Pb-Pb collisions can be enhanced and 
become larger than the value in pp collisions $N_{pp}^{X(3872)}/N_{pp}^{\psi(2S)}\simeq 0.1$ 
when treating X(3872) as a tightly bound state. Otherwise, the yield ratio is suppressed 
if X(3872) is a weakly bound state.  
Different degrees of charm quark kinetic thermalization are
studied. It is nonnegligible in X(3872) production,
which demonstrates the necessity
of realistic heavy quark evolution
in the study of X(3872) in heavy-ion collisions.
The coherent treatment of charm quark and D meson evolution in a
hot medium and the coalescence
process are necessary and meaningful
for studies on exotic candidates in heavy-ion collisions.

\vspace{0.5cm}
{\bf Acknowledgement:}
This work is supported by the National Natural Science Foundation of China
(NSFC) under Grant Nos. 11705125, 12047535 and 11975165.

%\end{spacing}
\end{document}